%% file: main.tex
\documentclass[10pt,twocolumn]{article}
\usepackage[english]{babel}
\usepackage[backend=biber,sorting=none,style=numeric-comp]{biblatex}
\usepackage[margin=2cm]{geometry} 
\usepackage{amsmath,amssymb}
\usepackage[version=4]{mhchem}
\usepackage{graphicx}
\usepackage{csquotes}
\usepackage{booktabs}
\usepackage{braket}
\usepackage{caption}
\usepackage{siunitx}
\usepackage{abstract}
\usepackage{subcaption}
\usepackage[acronym]{glossaries}
\usepackage[hidelinks]{hyperref}
\usepackage[capitalise]{cleveref}
\usepackage{verbatim}

\newacronym{NISQ}{NISQ}{Noisy Intermediate-Scale Quantum}
\newacronym{QAE}{QAE}{Quantum Autoencoders}
\newacronym{QGAN}{QGAN}{Quantum Generative Adversarial Networks}
\newacronym{GNN}{GNN}{Graph Neural Network}
\newacronym{VAE}{VAE}{Variational Autoencoder}
\newacronym{SNR}{SNR}{Signal-to-Noise Ratio}

\DeclareSIUnit\angstrom{\text{\AA}}
\DeclareSIUnit\calorie{cal}

\addto\captionsenglish{}

\title{Sparse Quantum Voxel Encoding for Readout-Efficient \\ Molecular Geometry Reconstruction on NISQ Devices}

\author{
Eros De Simone\textsuperscript{1,*}, Giuseppe Bifulco\textsuperscript{1}, Lorenza Di Mauro\textsuperscript{1},\\
Antonio Policicchio\textsuperscript{1}, Raoul Heese\textsuperscript{2}\\[6pt]
\small \textsuperscript{1}NTT DATA Italy \qquad \textsuperscript{2}NTT DATA Germany\\[4pt]
\small*corresponding author: eros.desimone@nttdata.com 
}
\date{\today}
\begin{document}
\twocolumn[
\maketitle
\vspace{-1cm}
\begin{abstract}
We propose a sparse computational-basis encoding of voxelized molecular geometries that converts molecular reconstruction from full-state tomography into support recovery by computational-basis sampling. To realize the encoding scheme, the molecular space is discretized into a 3D grid, and each atom's position and chemical species is mapped to a single computational basis state. This discretization introduces spatial quantization at the voxel-resolution scale. The molecule is then encoded as an equal superposition over this sparse set of occupied states, where we assume that a suitable state preparation method exists. In contrast to full state tomography, which requires on the order of $\mathcal{O}(3^n \times 10^{2\text{--}3})$ measurement shots, where $n$ is the number of qubits, our proposed encoding scheme reduces to a coupon-collector sampling problem in the computational basis. Complete recovery of an $A$-atom molecule requires $\mathcal{O}(A\log A)$ shots on noise-free hardware. On noisy hardware, the required number of shots increases. We demonstrate the method on the 156-qubit IBM Kingston device using 8-qubit circuits to reconstruct the discretized geometry of a 10-atom ethylamine molecule with high mean reconstruction recall using only $\mathcal{O}(10^2)$ shots despite substantial hardware noise. These results demonstrate that our proposed encoding scheme is a practical, readout-efficient representation for molecular geometries on near-term devices.
%
\vspace{1.cm}
\end{abstract}
]


\input{body}

\appendix
\crefalias{section}{appendix}
\input{appendix}

\printbibliography
\end{document}

%% file: body.tex
\section{Introduction} \label{sec:introduction}

The discovery and design of novel molecules with desired properties represents one of the most impactful challenges in computational chemistry and promotes many applications from therapeutic drugs to sustainable materials. However, the vast chemical space~\cite{polishchuk2013estimation} makes exhaustive exploration infeasible, motivating the development of an automated design of chemically valid structures, optimized for target properties such as binding affinity, solubility, or synthetic accessibility.

Machine learning has achieved remarkable success in this domain. \Glspl{GNN} learn molecular representations directly from connectivity patterns~\cite{gilmer2017neural,duvenaud2015convolutional}, \glspl{VAE} generate molecules by sampling from learned latent distributions~\cite{gomez2018automatic,kusner2017grammar}, and diffusion models have recently demonstrated state-of-the-art performance in de novo drug design~\cite{hoogeboom2022equivariant,xu2023geometric}.

In parallel, quantum computing~\cite{nielsen2010} is emerging as a promising new pillar for advances in chemistry, potentially capable of addressing problems that remain intractable for classical methods. While quantum chemistry traditionally focuses on simulating molecular Hamiltonians~\cite{cao2019quantum,mcardle2020quantum}, quantum computing also opens opportunities in machine learning~\cite{rodrigesdiaz2025}. In particular, by suitably encoding molecular properties into quantum states, quantum algorithms have the potential to explore and transform extremely large molecular libraries more efficiently than classical methods~\cite{zhao2026}. One prospectively promising application in this context is quantum generative modeling, where the aim is to sample molecular structures from a learned distribution while enforcing chemical validity and other desired constraints. However, resource-efficient molecular encoding is not straightforward and poses a fundamental challenge. Quantum variational autoencoders, Born machines, and quantum generative adversarial networks have been proposed for molecule generation~\cite{romero2017quantum,benedetti2019generative,wu2024qvae}, yet these models typically operate either in latent spaces or over indirect molecular descriptors, with generated samples requiring substantial classical post-processing.

Most existing quantum molecular encoding approaches face a fundamental measurement bottleneck: they do not permit direct, efficient readout of full molecular geometries from measurement outcomes. The usual amplitude encoding methods~\cite{wu2024qvae} embed structural features into probability amplitudes, achieving exponential information density but requiring full quantum state tomography to extract this information. Standard tomography requires measuring $3^n$ Pauli operator bases for an $n$-qubit system (corresponding to the tensor product of the three Pauli matrices across $n$ qubits), each estimated with $\mathcal{O}(10^{2\text{--}3})$ repetitions for sufficient statistical accuracy, yielding a total measurement cost scaling as $\mathcal{O}(3^n \times 10^{2\text{--}3})$~\cite{nielsen2010,Stricker_2022}. Even with recent advances in measurement-efficient tomography, such as threshold-based protocols that reduce the number of required measurement settings by up to two orders of magnitude~\cite{Zambrano2024tQST}, the total shot cost remains in the regime $\mathcal{O}(10^{4\text{--}5})$ for systems of moderate size, as each setting still requires $\mathcal{O}(10^{2\text{--}3})$ repetitions. Rotation encoding approaches~\cite{Pan:2025mmv} avoid the exponential preparation cost but suffer from reconstruction ambiguity: as noted in prior work~\cite{wu2024qvae}, ``it becomes intractable to reconstruct the input angles from the entangled quantum state.'' These trade-offs between encoding efficiency, expressivity, and measurement complexity limit the applicability of existing methods to structure-level generative tasks on \gls{NISQ} hardware, characterized by quantum processors containing up to a few hundred physical qubits which are not advanced enough yet for fault-tolerance or large enough to achieve a quantum advantage~\cite{Bharti_2022, preskill2018}.

In this work, we propose \emph{voxelization encoding}, a sparse computational-basis representation of molecular geometry that circumvents the measurement complexity problem through a measurement-efficient readout. By discretizing molecular space into a 3D grid and encoding each atom as a computational basis state representing both voxel position and atom type, we construct the molecular quantum state as an equal superposition. Within this work, we assume that a suitable state preparation method exists and focus on the decoding process. The key insight is that reconstructing this molecule does not require full state tomography. Instead, we exploit the statistical properties of equal superposition: for a molecule with $A$ atoms, measuring the quantum state repeatedly in the computational basis samples uniformly from the $A$ occupied basis states with probability $1/A$ each. The reconstruction problem thus reduces to the classical \emph{coupon collector problem}: observing all $A$ distinct coupons (atoms) through random sampling requires only $\mathcal{O}(A \log A)$ measurements. This transforms an exponentially expensive tomographic task into a statistically tractable sampling problem, enabling practical molecular encoding with $\mathcal{O}(10^{2\text{--}3})$ shots instead of the $\mathcal{O}(10^{4\text{--}6})$ total measurements required by full quantum state tomography on near-term quantum devices. We validate our proposed encoding scheme experimentally on quantum hardware. Our contribution is a \emph{readout-compatible} quantum representation, in which molecular geometry is laid out in computational-basis labels so that the structure can be recovered from raw measurements with only $\mathcal{O}(A \log A)$ shots. This is relevant, for example, when the output of a trained quantum model (e.g., a quantum generative model), where the result is inherently a quantum state and extracting molecular geometry would otherwise demand costly tomography. The disadvantage of the encoding approach is a loss in spatial precision, as each atom can only be localized up to the voxel grid coordinates.

In summary, our contributions are:
\begin{enumerate}
\item  We formalize quantum voxelization encoding as a sparse computational-basis support representation and prove that, once the corresponding support state is prepared, complete readout reconstruction of an $A$-atom molecule requires $\mathcal{O}(A \log A)$ shots with high confidence in the ideal noise-free setting.
\item We characterize the resolution-resource trade-off by deriving qubit requirements
and voxel-induced quantization errors, making explicit how the discretization, atom-type coverage, and quantum register sizes interact.
\item We demonstrate mean reconstruction recall (the fraction of atoms recovered at least once) $\bar{R}=0.94$ at $S=116$ shots and $\bar{R}=0.98$ at $S=200$ shots on \textit{IBM Kingston} for a representative 10-atom molecule (ethylamine, C$_2$H$_7$N), achieving two to three orders of magnitude reduction in measurement cost at the readout stage compared to amplitude encoding approaches requiring full state tomography.
\item We achieve molecule reconstruction on \textit{IBM Kingston} despite 74\% wasted-shot fraction from empty voxel detections and invalid quantum state indices, and compare against a representative noisy simulator running on classical hardware.
\end{enumerate}
To the best of our knowledge, this represents the first \gls{NISQ} demonstration of support recovery for a voxelized molecular atom-type point cloud, i.e., recovering atom positions and chemical species (without bonds, spin state, or other electronic-structure information), demonstrating practical feasibility. Hence, voxelization encoding may serve as a useful building block for quantum molecular generative modeling.

The remaining part of this manuscript is structured as follows. In \cref{sec:previous_work}, we provide a summary of related work. Our proposed encoding is presented in \cref{sec:voxelization}, the corresponding experimental results can be found in \cref{sec:experiments}. Finally, we present a conclusion in \cref{sec:conclusion}.

\section{Related Work} \label{sec:previous_work}

In analogy to a classical machine learning model, a quantum machine learning model also transforms input data to output data, where the data may either be classical or quantum. Each \emph{quantum model architecture} relies on a certain \emph{encoding scheme}, which determines how the data is represented on a quantum device~\cite{schuld2021machine, larose2020robust,weigold2021data,nielsen2010}. The choice of encoding also determines the measurement complexity: while some encodings permit efficient readout of encoded information, others (particularly those leveraging exponential Hilbert space capacity) may require exponentially costly tomographic reconstruction. Common encoding strategies include basis encoding (mapping discrete features directly to computational basis states), amplitude encoding (embedding data into probability amplitudes), angle encoding (mapping features to rotation angles), and kernel-based encodings (implicitly mapping data through quantum feature maps)~\cite{schuld2021machine, larose2020robust, weigold2021data, nielsen2010}:

\begin{enumerate}
\item Basis encoding represents classical data by mapping discrete features to computational basis states $\ket{x}$, where the integer $x$ directly encodes the input~\cite{nielsen2010, schuld2021machine}. This approach enables direct readout from measurements in the computational basis but is typically limited to discrete or digitized inputs. Our voxelization encoding can be considered as a form of basis encoding: each atom $a$ is assigned a unique integer index $c_a$ combining voxel position and atom type, which is then mapped to the computational basis state $\ket{c_a}$. The novelty lies in constructing an equal superposition over all such basis states, enabling simultaneous representation of the full molecular structure while preserving measurement efficiency.
\item Amplitude encoding embeds molecular structural features into probability amplitudes of a quantum state $\ket{\psi} = \sum_{i=0}^{2^n-1} \alpha_i \ket{i}$, where each amplitude $\alpha_i$ encodes classical data normalized to unit norm. QVAE-Mol~\cite{wu2024qvae} encodes 3D coordinates and atom types in a continuous amplitude space, achieving exponential information density: an $n$-qubit state can represent $2^n$ complex amplitudes, storing molecular features at polynomial quantum cost. However, extracting this information requires quantum state tomography, a procedure that demands exponentially many measurements. For an $n$-qubit system representing a molecule, conventional readout strategies are based on full quantum state tomography, which requires measuring up to $3^n$ Pauli observables, each estimated with $\mathcal{O}(10^{2\text{--}3})$ shots for sufficient statistical accuracy, yielding a total measurement cost scaling as $\mathcal{O}(3^n \times 10^{2\text{--}3})$. 
\item Angle encoding maps molecular features to the rotation angles of single-qubit or multi-qubit gates, typically through parameterized circuits of the form $U(\theta) = \prod_j R_j(\theta_j)$, where $\theta_j$ are functions of classical features. The recent MolQAE architecture~\cite{Pan:2025mmv} maps SMILES~\cite{doi:10.1021/ci00057a005} token sequences to parameterized U3 gates $U3(\theta, \phi, \lambda)$, allowing shallow $\mathcal{O}(n)$ state preparation with only 3 rotation parameters per qubit. Despite avoiding the exponential preparation cost, this approach encodes only 3 parameters per qubit (limited information capacity compared to amplitude encoding's $2^n$ amplitudes), and suffers from reconstruction ambiguity. Without auxiliary quantum registers or classical post-processing constraints (e.g., enforcing physical symmetries), the inverse problem of recovering rotation angles from measurement statistics is ill-posed, particularly for highly entangled states where local measurements provide limited information about global parameters.
\item Kernel-based encoding maps classical data into a quantum feature space through parameterized circuits, defining an implicit feature map in Hilbert space~\cite{schuld2021machine}. Rather than extracting full quantum states, this approach computes similarity measures (kernels) between data points via inner product estimation, enabling classification and regression tasks through quantum support vector machines. Since only pairwise similarities are required rather than complete state reconstruction, kernel-based methods largely avoid the measurement bottleneck affecting amplitude and angle encodings. However, they are not directly applicable to molecular generative modeling, where the objective is to produce explicit 3D atomic geometries rather than to compute pairwise distances between molecular representations.
\end{enumerate}

Beyond encoding-specific challenges, efficient measurement strategies have been developed for specific observable classes. Classical shadows~\cite{huang2020predict} enable efficient estimation of expectation values $\langle O \rangle$ with polynomial overhead, achieving $\varepsilon$-accurate estimation with $\mathcal{O}(\log(M)/\varepsilon^2)$ measurements for $M$ observables, while commuting-Pauli grouping strategies~\cite{Verteletskyi_2020} reduce the number of required measurement settings. Neural quantum state tomography leverages machine learning to reduce measurement requirements for specific tasks~\cite{torlai2018neural}. However, these methods are fundamentally limited to linear observables and low-rank approximations; they cannot efficiently reconstruct complete 3D molecular geometries encoded in exponentially large state spaces. For generative modeling tasks requiring direct structural sampling, where the goal is to produce chemically valid 3D conformers rather than estimate average properties, these measurement strategies remain insufficient. The reconstruction of full molecular configurations from quantum states thus remains an open challenge for practical quantum molecular design on \gls{NISQ} hardware. Our work addresses the measurement bottleneck by proposing a voxelization encoding scheme that inherently enables measurement-efficient reconstruction.

\section{Quantum Voxelization Encoding} \label{sec:voxelization}

Voxelization encoding is a form of basis encoding~\cite{nielsen2010, schuld2021machine}: each atom is mapped to a computational basis state, enabling direct readout. While the resulting molecular state is technically a superposition, the encoded information resides entirely in \textit{which} basis states are occupied, not in the amplitude magnitudes. The amplitude magnitudes, in our case, only encode binary information (yes/no) about the presence of a specific atom, rather than encoding a property. We encode atomic positions and types directly in the computational basis labels of an equal superposition state, transforming reconstruction from exponential tomography into polynomial coupon collection. This approach is described in more detail in the following.

Consider a molecule
\begin{equation} \label{eq:M}
M:=\{(x_1,y_1,z_1,\tau_1),\dots,(x_A,y_A,z_A,\tau_A)\}
\end{equation}
consisting of $A$ atoms, where each atom $a$ is characterized by its 3D coordinates $(x_a, y_a, z_a) \in \mathbb{R}^3$ and its atom type $\tau_a \in \mathcal{T}$ (e.g., C, H, O, N). 
Our goal is to encode this molecular structure in a quantum state $\ket{\psi_{M^*}}$ such that: (i) the molecular information is stored in the computational-basis support of the state, (ii) measurement of the quantum state provides sufficient information to reconstruct the molecular structure with complete recall at voxel-level precision, and (iii) the encoding preserves spatial relationships and atom-type information necessary for downstream quantum machine learning tasks. We do not claim that arbitrary sparse molecular support
states of this form can be prepared efficiently in general since state preparation is a separate problem beyond the scope of this work.

\subsection{Quantum State Encoding}
\label{subsec:voxelization_framework}
We discretize the molecular coordinate space into a regular 3D grid $(i,j,k)$ with resolution $V$ along each dimension, creating $V^3$ voxels with an edge length of $s_{\text{voxel}}$. The molecular structure is first centered at its geometric centroid, and the resulting coordinates are shifted such that the voxel grid spans symmetrically around the origin. For each atom $a \in \{1,\dots,A\}$ with centered 3D coordinates $(x_a, y_a, z_a)$, the voxel coordinates $i_a, j_a, k_a \in \{0, 1, \ldots, V-1\}$ are computed according to
\begin{align}
i_a &:= \left\lfloor \frac{x_a + r_{\text{grid}}}{s_{\text{voxel}}} \right\rfloor, \\
j_a &:= \left\lfloor \frac{y_a + r_{\text{grid}}}{s_{\text{voxel}}} \right\rfloor, \\
k_a &:= \left\lfloor \frac{z_a + r_{\text{grid}}}{s_{\text{voxel}}} \right\rfloor,
\end{align}
where $r_{\text{grid}} := \frac{V \cdot s_{\text{voxel}}}{2}$ is the half-extent of the grid along each axis. The resulting voxel indices are clipped to the valid range $[0, V-1]$ to handle numerical edge cases. 

It is presumed that each voxel contains at most one atom. If multiple atoms map to the same voxel during discretization, the grid parameters must be adjusted until all atoms occupy distinct voxels, either by increasing the resolution $V$ or decreasing the voxel size $s_{\text{voxel}}$. For a given molecule, collisions can always be avoided by decreasing the voxel size: two atoms separated by at least the minimum inter-atomic distance $d_{\min}$ cannot share a voxel once its diagonal satisfies $\sqrt{3}\,s_{\text{voxel}} < d_{\min}$. A collision-free grid is therefore guaranteed and can be reached by iteratively decreasing $s_{\text{voxel}}$ (equivalently, increasing $V$) in a linear number of refinement steps.
To encode datasets containing multiple molecules, a universal grid configuration must be established that prevents collisions across all molecular structures, potentially through systematic exploration of molecular orientations and scaling factors. The formal algorithmic treatment of optimal collision-free grid design for molecular datasets is beyond the scope of this work and remains a possible direction for future investigation.

Presuming that a suitable voxel grid is found, each voxel $(i_a, j_a, k_a)$ is assigned to a linear index $v_a \in \{0, 1, \ldots, V^3-1\}$ using the standard row-major ordering
\begin{equation}
\label{eq:voxel_1d_mapping}
v_a := i_a \cdot V^2 + j_a \cdot V + k_a.
\end{equation}

In addition, the type of each atom also needs to be encoded, for which we employ a \textit{product encoding} strategy by combining the space of all voxel coordinates with the space of all atom types. This leads to a combined linear index. Let $\mathcal{T}$ denote the set of $T:=|\mathcal{T}|$ distinct atom types present in the molecule. The type of atom $a$ is given by $\tau_a \in \mathcal{T} := \{0,\dots,T-1\}$, where we use atom names and numbers interchangeably. The combined linear index for the position and type of atom $a$ then reads
\begin{equation}
\label{eq:combined_index}
c_a := v_a \cdot T + \tau_a
\end{equation}
with $c_a \in \mathcal{C}:=\{0, 1, \ldots, V^3T-1\}$. The set $\mathcal{C}$ represents all $C:=|\mathcal{C}|=V^3T$ possible configurations for a single atom in terms of its position (via voxel coordinates) and its type. A molecule with $A$ atoms consequently has $C^A$ possible configurations. As already mentioned above, we presume that different atoms do not share the same voxel coordinates. This renders some of the configurations infeasible, leaving 
\begin{equation}
\label{eq:hatC}
    \hat{C}:=\prod_{a=1}^A(C-(a-1)T)
\end{equation}
feasible molecule configurations (which contain only atoms at different voxel coordinates), as represented by the configuration set $\hat{\mathcal{C}}$ with $|\hat{\mathcal{C}}|=\hat{C}$. Additional combinatorial restrictions are presented in \cref{app:molconf}.

The voxelization encoding of molecule $M$, \cref{eq:M}, consequently reads
\begin{equation} \label{eq:Menc}
    M^*:=\{c_1,\dots,c_A\} \in \hat{\mathcal{C}}.
\end{equation}
The difference between $M$ and $M^*$ is that the latter has a lower spatial resolution of the atom positions according to the chosen voxel grid.

So far, this is a purely classical data processing approach. However, it enables a direct mapping onto a quantum system using basis encoding. To that end, each atom configuration $c \in M^*$ is mapped to a unique computational basis state of a quantum system $\ket{c} \in \mathcal{H}$. Since each $c \in \{0,\dots,C-1\}$ indexes a single-atom configuration, the required Hilbert space $\mathcal{H}$ has $n := \lceil \log_2 C \rceil = \lceil \log_2 (V^3 T) \rceil$ qubits in total. The molecule $M^*$ can then be encoded as an equal superposition of all basis states that represent its atom configurations
\begin{equation} \label{eq:quantum_state}
\ket{\psi_{M^*}} := \frac{1}{\sqrt{|M^*|}} \sum_{c_a \in M^*} \ket{c_a}.
\end{equation}
By construction, only the first $C$ basis states $\ket{0},\dots,\ket{C-1}$ can encode a valid single-atom configuration, whereas the last $2^n-C$ basis states are unused. The encoded quantum state $\ket{\psi_{M^*}} \in \mathcal{H}$ contains the full information about all atom positions and types. 

The construction in \cref{eq:quantum_state} should be understood as a readout-oriented representation, not as a general solution to quantum data loading. The measurement advantage derived below applies once a state with the desired sparse support has been prepared, or when such a state is produced as the output of a (generative) quantum model. However, preparing an arbitrary state $\ket{\psi_{M^*}}$ may in fact not scale efficiently. The contribution of the present work is therefore the decoding structure: molecular reconstruction is reduced from full-state tomography to support recovery by computational-basis sampling. In the experiments below, $\ket{\psi_{M^*}}$ is prepared using a generic isometry-based initialization routine~\cite{PhysRevA.93.032318}. This approach is suitable for a proof-of-concept demonstration but has limited scalability.


\subsection{Decoding via Measurement} \label{subsec:measurement_reconstruction}

To decode the quantum state $\ket{\psi_{M^*}}$ back into a molecule representation of the form of \cref{eq:M}, it can be repeatedly measured in the computational basis. 
After each measurement, the original quantum state $\ket{\psi_{M^*}}$ is destroyed and must be prepared again. The whole process of preparing the state and measuring the bit value of each qubit is known as a \emph{shot}.
Each shot effectively ``collapses'' $\ket{\psi_{M^*}}$, a superposition of basis states, into a single basis state $\ket{b}$. As a consequence, each qubit $q \in \{1,\dots,n\}$ attains a definite bit value $b_q \in \{0,1\}$, which allows to determine the outcome
\begin{equation} \label{eq:b}
b:=\sum_{q=1}^{n}2^{q-1}b_q \in \{0,\dots,2^n-1\}.
\end{equation}
This is an intrinsically non-deterministic process~\cite{tomaz2025} with the probability distribution
\begin{equation} \label{eq:pM}
p_{M^*}(b) := |\braket{b|\psi_{M^*}}|^2 = \frac{1}{|M^*|} \sum_{c_a \in M^*} \delta_{b,c_a}
\end{equation}
with support $M^*$.
In other words, each shot corresponds to drawing a sample $b \sim p_{M^*}(b)$, which represents one of the $|M^*|=A$ occupied basis states with uniform probability $1/A$, effectively revealing both the voxel position and the atom type of one randomly selected atom out of all atoms of the encoded molecule $M^*$.

The molecular reconstruction problem thus reduces to the classical \textit{coupon collector problem}~\cite{alma9910292682105765}: given $A$ distinct coupons (atoms), how many random draws (shots) are required to observe all coupons at least once?
For $A$ atoms, the expected number of measurements needed to observe all atoms $S$ is given by
\begin{equation} \label{eq:coupon_expected}
    \mathbb{E}[S] = A \sum_{i=1}^A \frac{1}{i} = A \cdot H_A \approx A (\ln A + \gamma),
\end{equation}
where $H_A$ is the $A$-th harmonic number and $\gamma \approx 0.577$ is the Euler-Mascheroni constant. This scales as $\mathcal{O}(A \log A)$. To achieve complete reconstruction with high confidence $1 - \epsilon$, the required number of shots is approximately
\begin{equation} \label{eq:shots_confidence}
    S_{\epsilon} \approx A \left( \ln A + \ln \frac{1}{\epsilon} \right).
\end{equation}
For example, to reconstruct a 10-atom molecule with 99\% confidence ($\epsilon = 0.01$), we require $S_{0.01} \approx 10(\ln 10 + \ln 100) \approx 70$ shots. This is substantially more efficient than amplitude encoding approaches with a generally exponential scaling of the required number measurements.

However, these estimates only consider an idealized situation, where the quantum computer is able to generate uniform samples in an accurate and unbiased way. Currently, this is only possible using a simulated quantum computer (running on classical hardware) since the hardware imperfections of \gls{NISQ} devices only allow noisy sampling processes.
This may lead to the detection of atom configurations that are not part of the encoded molecule (i.e., $b \not\in M^*$) or not even of the encoding space (i.e., $b \geq C$). Conversely, other encoded atom configurations may remain undetected even with a high number of shots.

\subsection{Quantization Error} \label{subsec:quantization_error}

The voxelization map replaces each continuous atomic coordinate by a discrete voxel index such that sub-voxel spatial information is lost. When the encoded molecule $M^*$ is decoded back to a continuous representation of the form of \cref{eq:M}, each atom can for example be assigned to the center of its voxel,
\begin{equation}
\label{eq:voxel_center}
    \hat{x}_a := \left(i_a + \tfrac{1}{2}\right) s_{\text{voxel}} - r_{\text{grid}},
\end{equation}
and analogously for $\hat{y}_a$ and $\hat{z}_a$ using $j_a$ and $k_a$, respectively. This is the estimator minimizing the worst-case error within a voxel. Since the true coordinate lies somewhere inside the voxel cube of edge length $s_{\text{voxel}}$, the error along each axis is bounded by $s_{\text{voxel}}/2$. The maximum Euclidean coordinate error per atom is therefore attained at a voxel corner and bounded by the half-diagonal of the voxel cube: $\Delta_{\max} := \frac{\sqrt{3}}{2} s_{\text{voxel}}$. Assuming atomic positions to be uniformly distributed within their voxels, the corresponding root-mean-square error is $\Delta_{\text{rms}} := s_{\text{voxel}}/2$. For the grid used in our experiments ($s_{\text{voxel}} = \SI{0.95}{\angstrom}$), this yields a worst-case error $\Delta_{\max} \approx \SI{0.82}{\angstrom}$ and an RMS error $\Delta_{\text{rms}} \approx \SI{0.48}{\angstrom}$. This quantization error sets a fundamental resolution limit of the encoding: it can only be reduced by decreasing $s_{\text{voxel}}$ at the cost of a larger encoding space and more qubits. The resulting trade-off between spatial resolution and qubit count is intrinsic to voxelization and is discussed further in \cref{sec:conclusion}.

\section{Experimental Validation} \label{sec:experiments}

\begin{figure*}[ht!]
\centering
\includegraphics[width=\columnwidth]{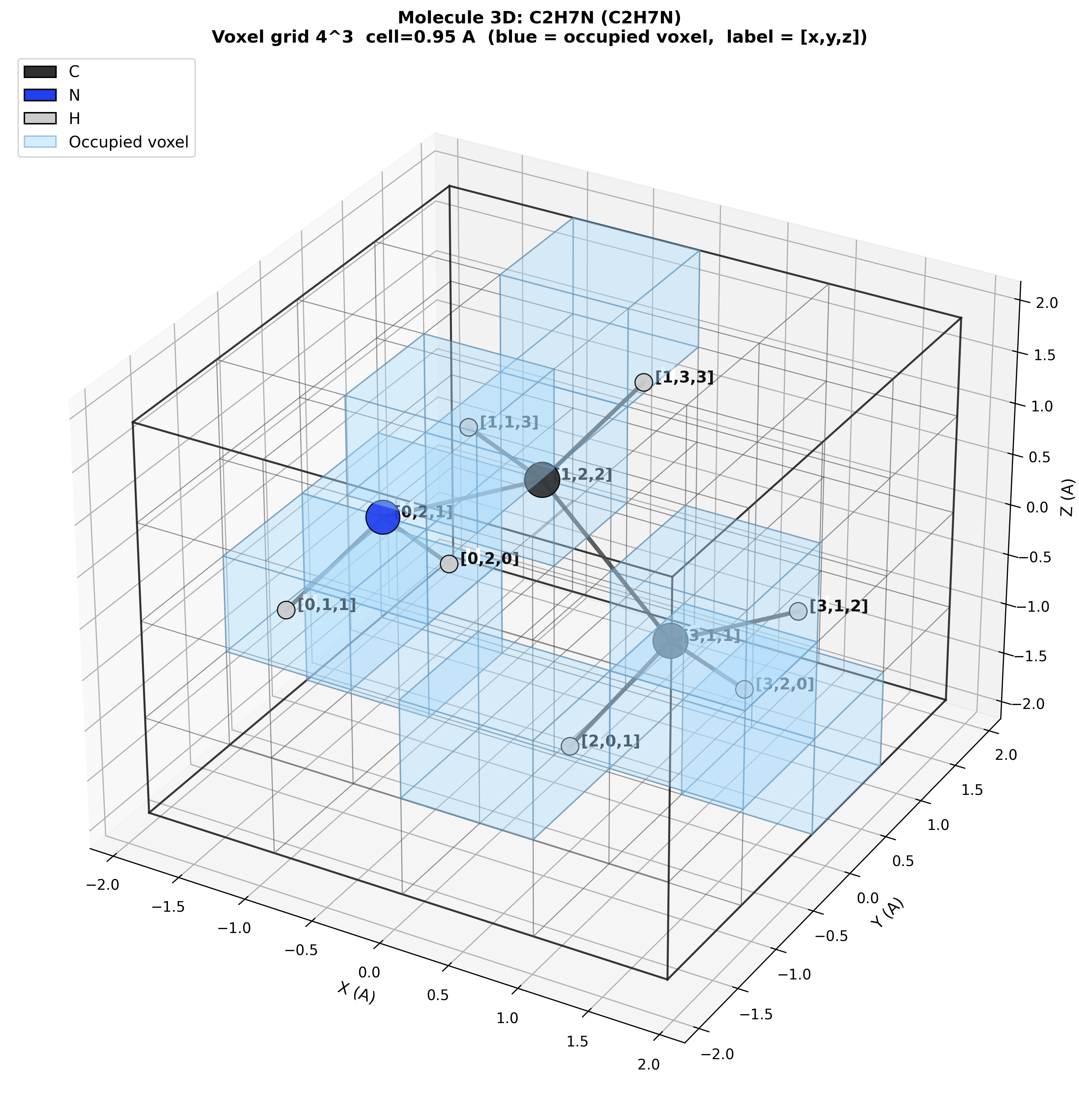}
\caption{Ethylamine (\ce{C2H7N}) encoded in a $4\times4\times4$ voxel grid ($s_\text{voxel}=\SI{0.95}{\angstrom}$).
Carbon atoms (dark gray), nitrogen (blue), and hydrogen (light gray) are shown at their MMFF-optimized 3D positions after centering.
Each atom is labeled with the corresponding voxel indices.
Shaded voxel boxes highlight the cells containing atoms (e.g., N is in the voxel $(0,2,1)$).}
\label{fig:molecule_3d}
\end{figure*}

As a proof of concept of the voxelization encoding framework, we focus on the encoding of a single molecule, ethylamine (\ce{C2H7N}), a representative 10-atom organic molecule. We use the preprocessing approach from \cref{app:preprocessing} to obtain the 3D atom positions for the ethylamine molecule and consider $T = 3$ atom types $\mathcal{T} := \{\text{C, H, N}\}$ for the encoding. Based on the molecular geometry, we choose a grid resolution of $V := 4$ along each spatial dimension, which results in $V^3 = 64$ voxels. The voxel edge length $s_{\text{voxel}} := \SI{0.95}{\angstrom}$ is smaller than the minimum inter-atomic distance in the optimized molecular geometry (\SI{1.09}{\angstrom} for \ce{C-H} bonds). Note that the worst-case collision-free condition $\sqrt{3}\,s_{\text{voxel}} < d_{\min}$ (\cref{subsec:voxelization_framework}) would require $s_{\text{voxel}} < \SI{0.63}{\angstrom}$, hence a finer grid and more qubits. We instead adopt the coarser grid with $s_{\text{voxel}} = \SI{0.95}{\angstrom}$, which is collision-free for this molecule under the canonical orientation and centering applied during preprocessing, as verified during encoding (all atoms occupy distinct voxels, see \cref{tab:encoding}). The resulting encoding space has dimension $C = V^3 T = 192$, requiring $n = \lceil \log_2(192) \rceil = 8$ qubits, with 10 atoms occupying this 192-dimensional space. The complete molecule data is listed in \cref{tab:encoding} and, additionally, visualized in \cref{fig:molecule_3d}.

\begin{table*}[ht!]
\centering
\setlength{\tabcolsep}{5pt}
\begin{tabular}{ccrrrcccccr}
\toprule
Atom & Type & $x$ [\si{\angstrom}] & $y$ [\si{\angstrom}] & $z$ [\si{\angstrom}] 
     & $i_a$ & $j_a$ & $k_a$ & $v_a$ & $\tau_a$ & $c_a$ \\
\midrule
0 & C &  0.991 & -0.248 & -0.177 & 3 & 1 & 1 & 53 & 0 & 159 \\
1 & C & -0.215 &  0.223 &  0.618 & 1 & 2 & 2 & 26 & 0 &  78 \\
2 & N & -1.420 &  0.222 & -0.203 & 0 & 2 & 1 &  9 & 2 &  29 \\
3 & H &  1.179 &  0.405 & -1.036 & 3 & 2 & 0 & 56 & 1 & 169 \\
4 & H &  1.885 & -0.241 &  0.454 & 3 & 1 & 2 & 54 & 1 & 163 \\
5 & H &  0.848 & -1.269 & -0.548 & 2 & 0 & 1 & 33 & 1 & 100 \\
6 & H & -0.368 & -0.430 &  1.483 & 1 & 1 & 3 & 23 & 1 &  70 \\
7 & H & -0.039 &  1.234 &  0.997 & 1 & 3 & 3 & 31 & 1 &  94 \\
8 & H & -1.583 & -0.715 & -0.570 & 0 & 1 & 1 &  5 & 1 &  16 \\
9 & H & -1.279 &  0.819 & -1.017 & 0 & 2 & 0 &  8 & 1 &  25 \\
\bottomrule
\end{tabular}
\caption{Voxelization encoding of ethylamine (\ce{C2H7N}). 
Coordinates $(x,y,z)$ are given after centering at the molecular centroid (in \si{\angstrom}, no scaling applied).
Voxel indices $(i,j,k)$ are computed with grid resolution $V=4$ and voxel size $s_\text{voxel}=\SI{0.95}{\angstrom}$.
The linear voxel index is $v_a = i\cdot V^2 + j\cdot V + k$, the atom-type index is
$\tau_a \in \{0\,(\text{C}),\,1\,(\text{H}),\,2\,(\text{N})\}$, and the combined
index $c_a = v_a \cdot T + \tau_a$ (with $T=3$) identifies the computational basis state $|c_a\rangle$
used in the quantum encoding, \cref{eq:quantum_state}.}
\label{tab:encoding}
\end{table*}

The number of shots is inferred from the coupon collector analysis from \cref{subsec:measurement_reconstruction}. For ethylamine with $A=10$ atoms, we execute experiments with $S \in \{47, 70, 116, 200, 300\}$ shots, spanning the theoretically predicted range from 90\% confidence ($S=47$ shots) to beyond 99.99\% confidence ($S=116$ shots) as calculated in the ideal noiseless case; see \cref{app:shots_calculation}. Furthermore, we also consider $S=200$ and $S=300$ to take noisy effects into account. We perform $N=10$ independent experiments (repetitions) per shot count to enable statistical characterization of recall variance and noise fluctuations, yielding a total of 50 quantum circuit executions and \num{7330} total quantum measurements.

\subsection{Implementation and Hardware} \label{subsec:implementation_hardware}

We implement the voxelization encoding framework in Qiskit~\cite{qiskit} and execute experiments on \textit{IBM Kingston}, a 156-qubit superconducting transmon processor. Device characteristics including coherence times, gate errors, and readout fidelities are reported in~\cref{app:device_properties}. All hardware experiments were executed in April 2026 via the IBM Quantum Platform. For testing purposes and noise characterization, we also employ a noisy simulator, which runs on classical hardware. It emulates device noise from stored calibration data of \textit{IBM Fez}, another 156-qubit IBM processor sharing the same heavy-hexagon topology as \textit{IBM Kingston}, which serves as a representative noisy backend of comparable architecture since a noisy simulator based on the calibration data of \textit{IBM Kingston} was not available at the time of our experiments.

We employ Qiskit's \texttt{initialize} method for state preparation via isometry decomposition~\cite{PhysRevA.93.032318}, which synthesizes a quantum circuit implementing the unitary transformation $U\ket{0}^{\otimes n} = \ket{\psi_{M^*}}$.  Full details of the transpilation procedure and the resulting circuit depth distribution are reported in~\cref{app:transpilation}. No error mitigation techniques were applied to preserve raw hardware measurement statistics and assess intrinsic noise resilience of the voxelization encoding.

\subsection{Measurement Statistics} \label{subsec:measurement_statistics}

\begin{figure}[ht!]
\centering
\begin{subfigure}{\columnwidth}
\includegraphics[width=\columnwidth]{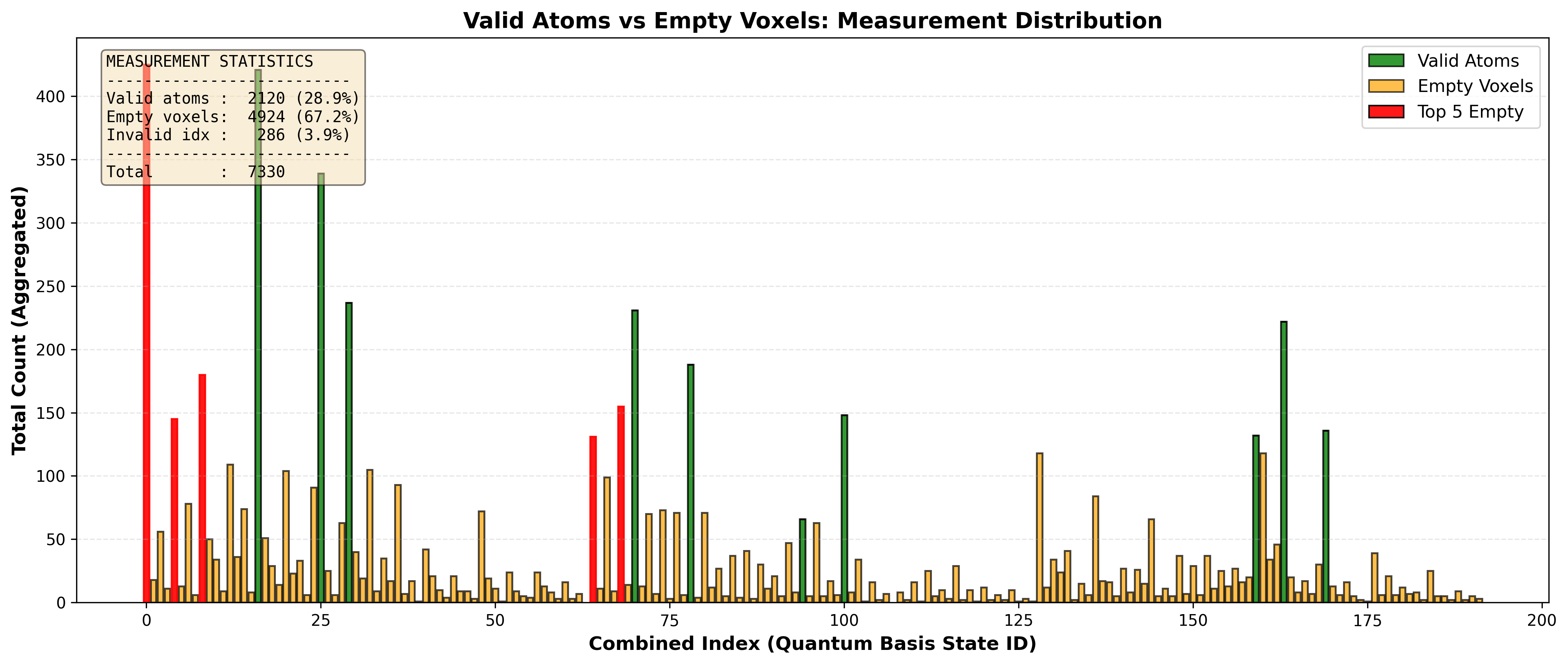}
\caption{\emph{IBM Kingston}}
\label{fig:valid_vs_empty:hw}
\end{subfigure}
\\[.2cm]
\begin{subfigure}{\columnwidth}
\includegraphics[width=\columnwidth]{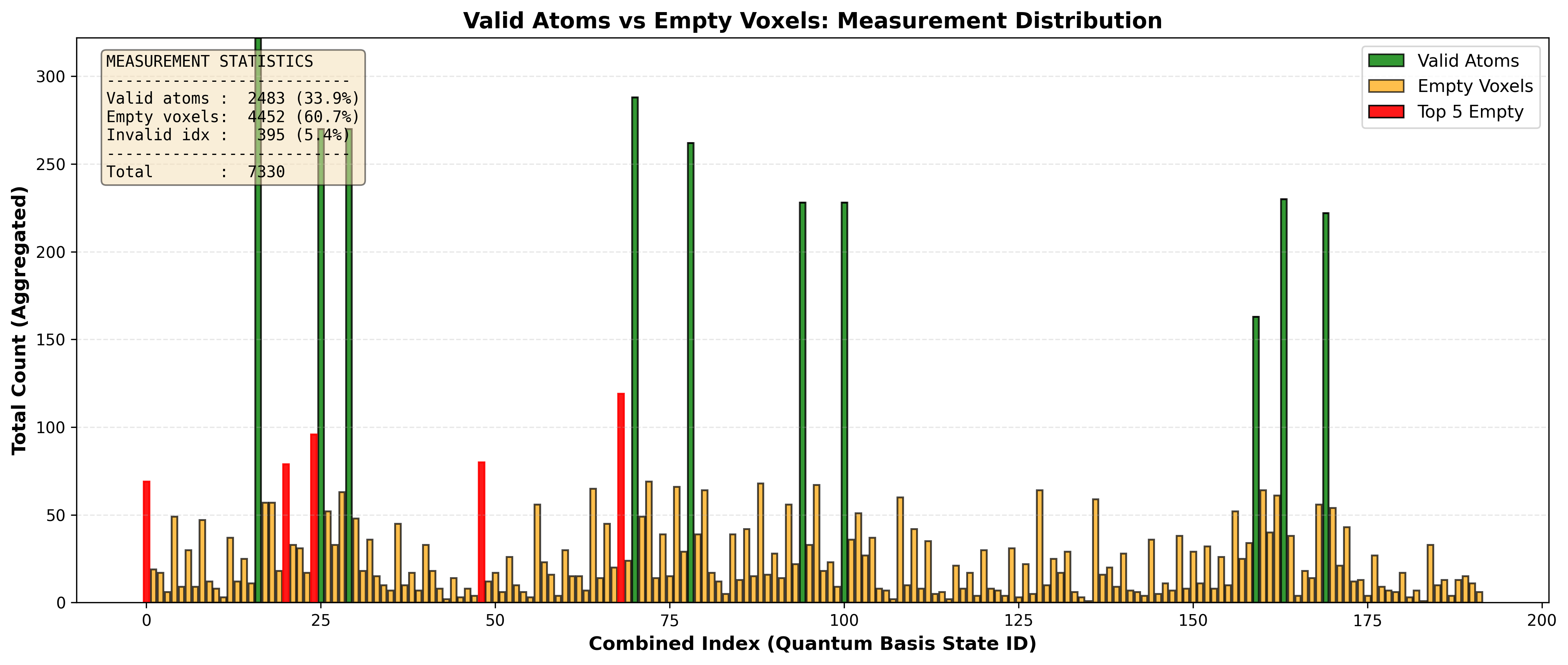}
\caption{Noisy simulator}
\label{fig:valid_vs_empty:sim}
\end{subfigure}
\caption{Distribution of measurement outcomes. Green bars indicate the 10 valid atomic positions, orange bars represent empty voxel detections across 182 unoccupied grid indices. Red highlights mark the 5 most frequently observed empty voxels.}
\label{fig:valid_vs_empty}
\end{figure}

To begin with, we present the measurement outcome distributions across the $C=192$ computational basis states in \cref{fig:valid_vs_empty}. The outcome $b$, \cref{eq:b}, for each shot is classified as
\begin{itemize}
    \item \emph{correct atom}, if $b \in M^*$ (valid molecular feature),
    \item \emph{empty voxel}, if $0 \leq b < C \,\land\, b \notin M^*$ (unoccupied position), or
    \item \emph{invalid index}, if $b \geq C$ (outside encoding subspace).
\end{itemize}
The hardware distribution in \cref{fig:valid_vs_empty:hw} exhibits striking bias toward low-index states, with pronounced peaks at $b=0$ (ground state $|00\cdots0\rangle$). We presume that this behavior is a result of progressive decoherence: as $T_1$ relaxation acts, probability amplitude preferentially leaks toward the ground state, a well-documented effect in superconducting qubits~\cite{Krantz_2019}. The simulator, on the other hand, exhibits approximately uniform error distribution across all 182 states not encoding atoms, reflecting its idealized depolarizing noise model with symmetric Pauli errors. The absence of progressive decoherence prevents preferential ground-state leakage, democratically distributing errors across unoccupied voxel space. However, a detailed discussion of the noise model of the simulator and its deviations from the hardware are beyond the scope of this work.

\begin{figure}[ht!]
\centering
\begin{subfigure}{\columnwidth}
\includegraphics[width=\columnwidth]{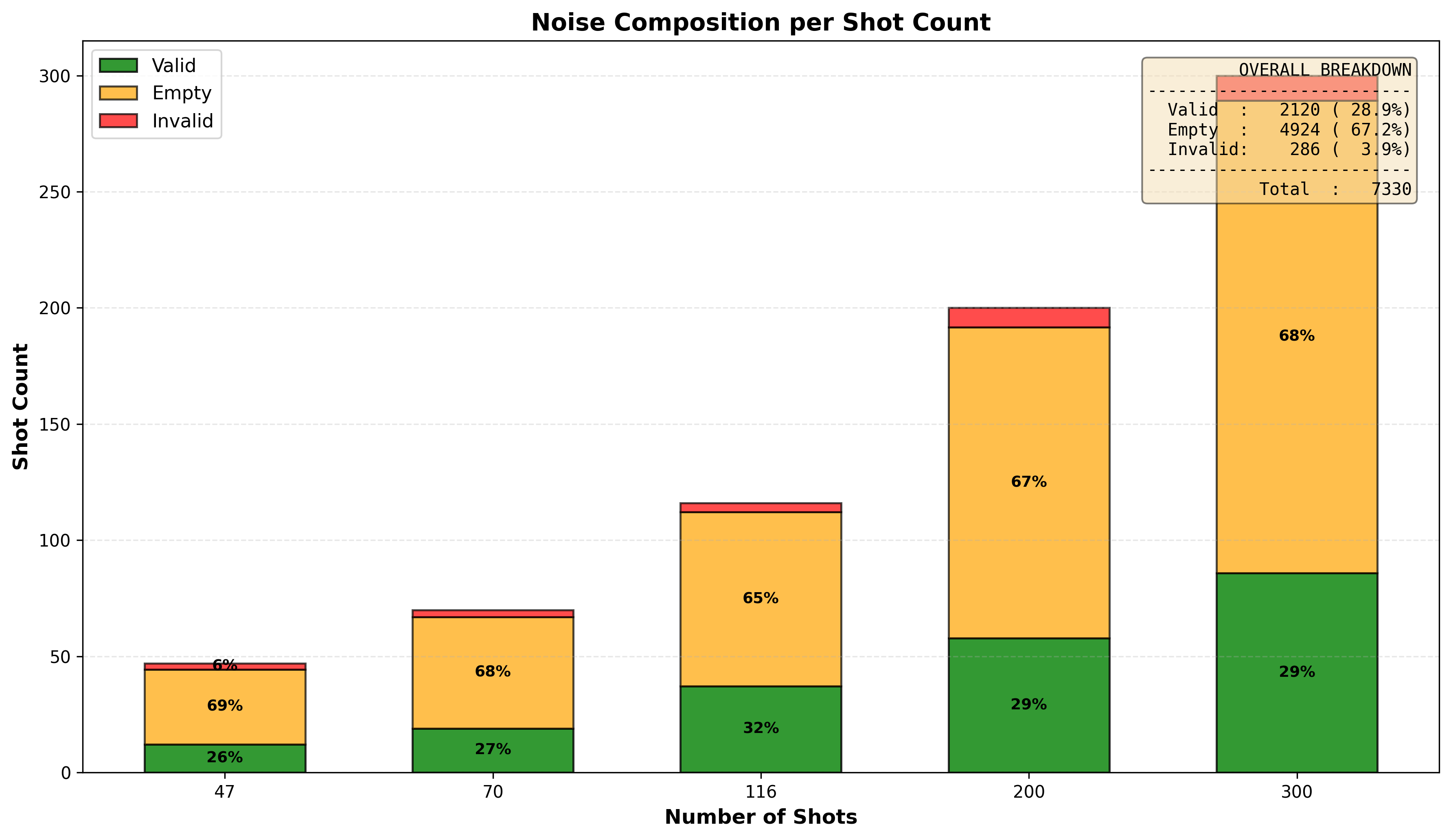}
\caption{\emph{IBM Kingston}}
\label{fig:noise_composition:hw}
\end{subfigure}
\\[.2cm]
\begin{subfigure}{\columnwidth}
\includegraphics[width=\columnwidth]{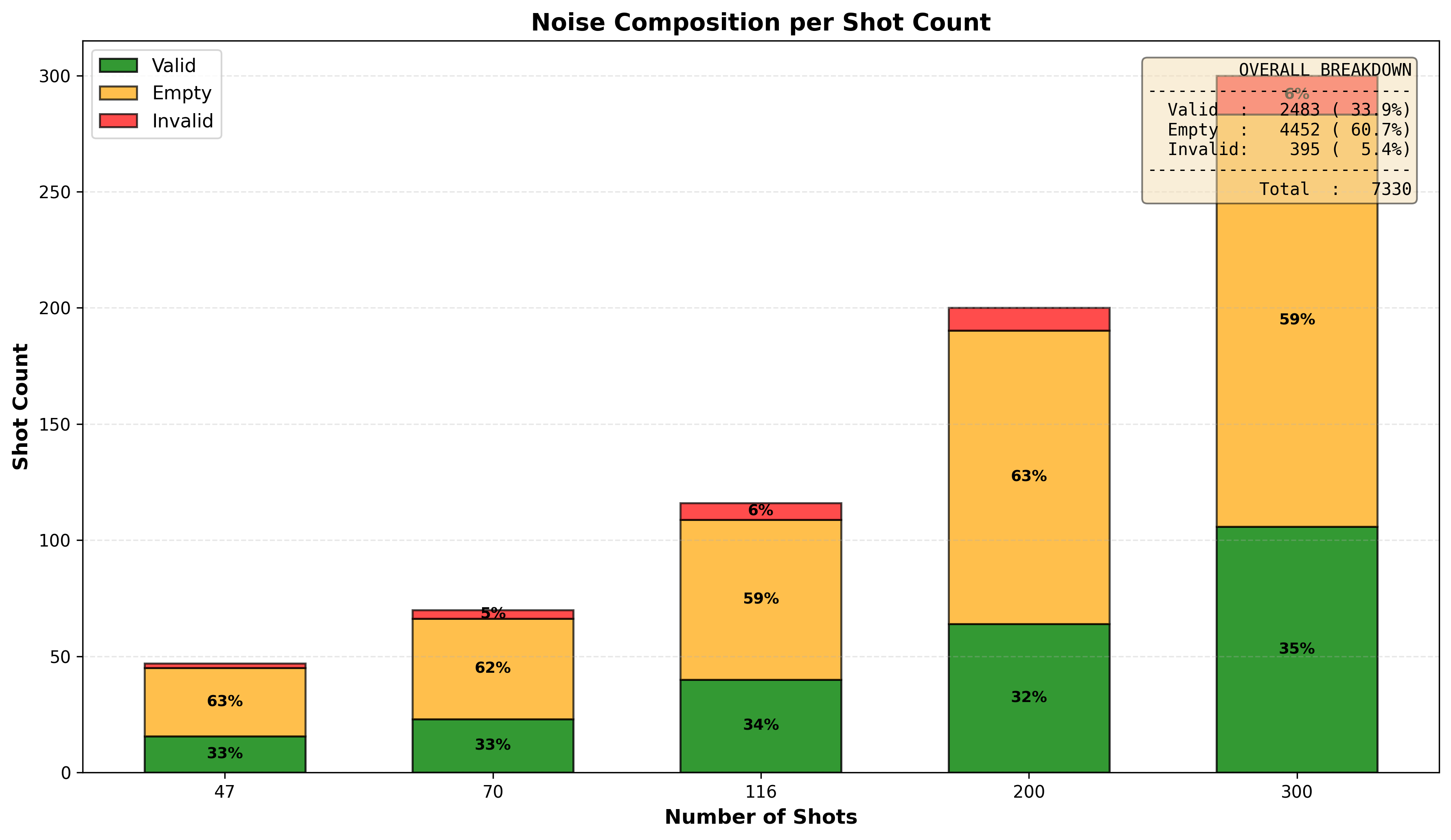}
\caption{Noisy simulator}
\label{fig:noise_composition:sim}
\end{subfigure}
\caption{Noise composition breakdown across shot counts (47--300). Stacked bars decompose measurements into three categories: valid atomic detections (green), empty voxel errors (orange), and invalid quantum state indices beyond the encoding space (red).}
\label{fig:noise_composition}
\end{figure}

Complementary to the outcome distribution, we also present the aggregated outcome classifications across all shots in \cref{fig:noise_composition}. We find that the overall noise breakdown on hardware reveals 28.9\% valid measurements, 67.2\% empty voxels, and 3.9\% invalid indices. On the simulator, the breakdown is 33.9\% valid, 60.7\% empty voxels, and 5.4\% invalid indices. Notably, the noise composition remains remarkably stable across the shot count range (47--300) on both hardware and simulator, indicating that noise characteristics are not significantly affected by measurement budget within this regime.

\begin{figure}[ht!]
\centering
\begin{subfigure}{\columnwidth}
\includegraphics[width=\columnwidth]{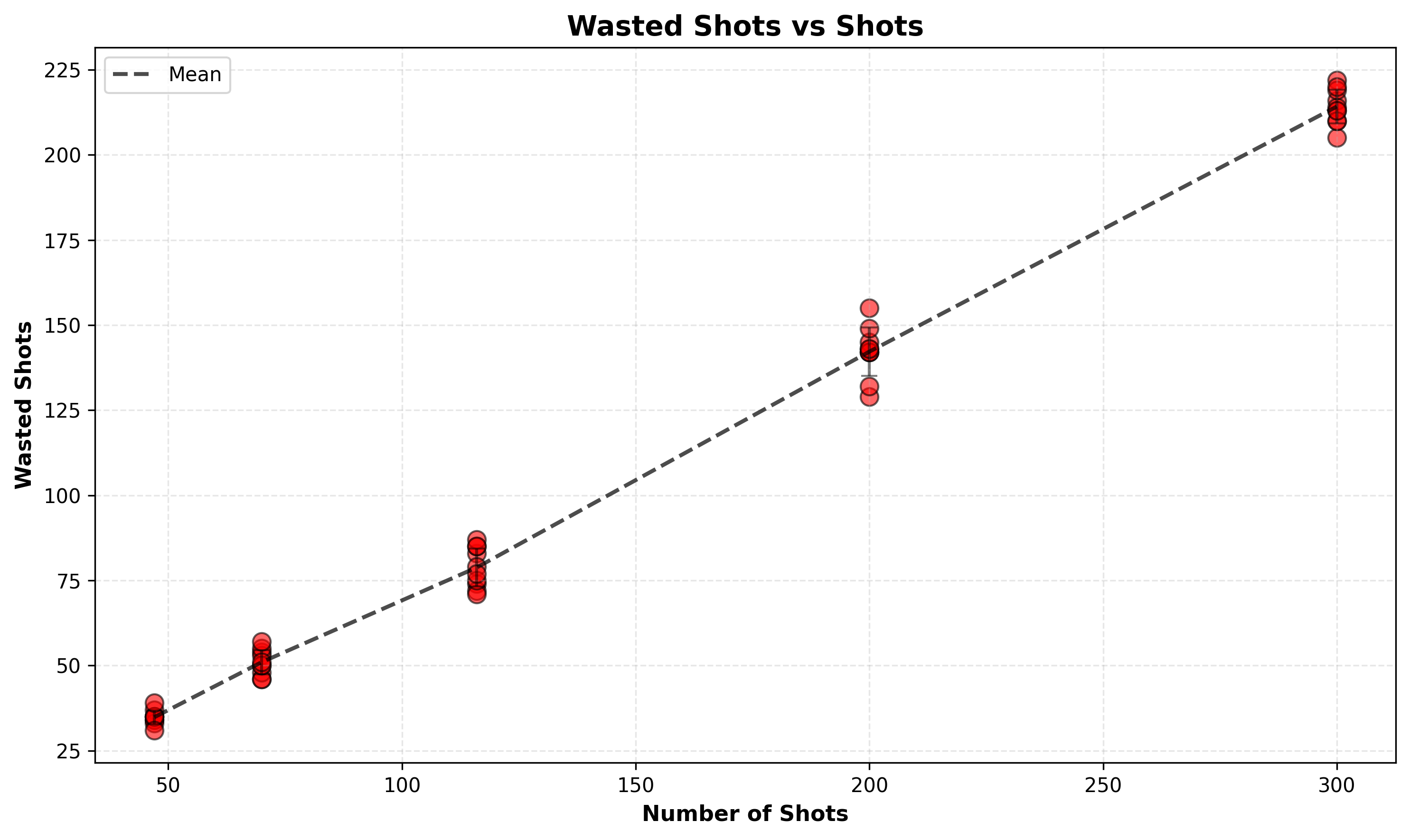}
\caption{\emph{IBM Kingston}}
\label{fig:wasted_vs_shots:hw}
\end{subfigure}
\\[.2cm]
\begin{subfigure}{\columnwidth}
\includegraphics[width=\columnwidth]{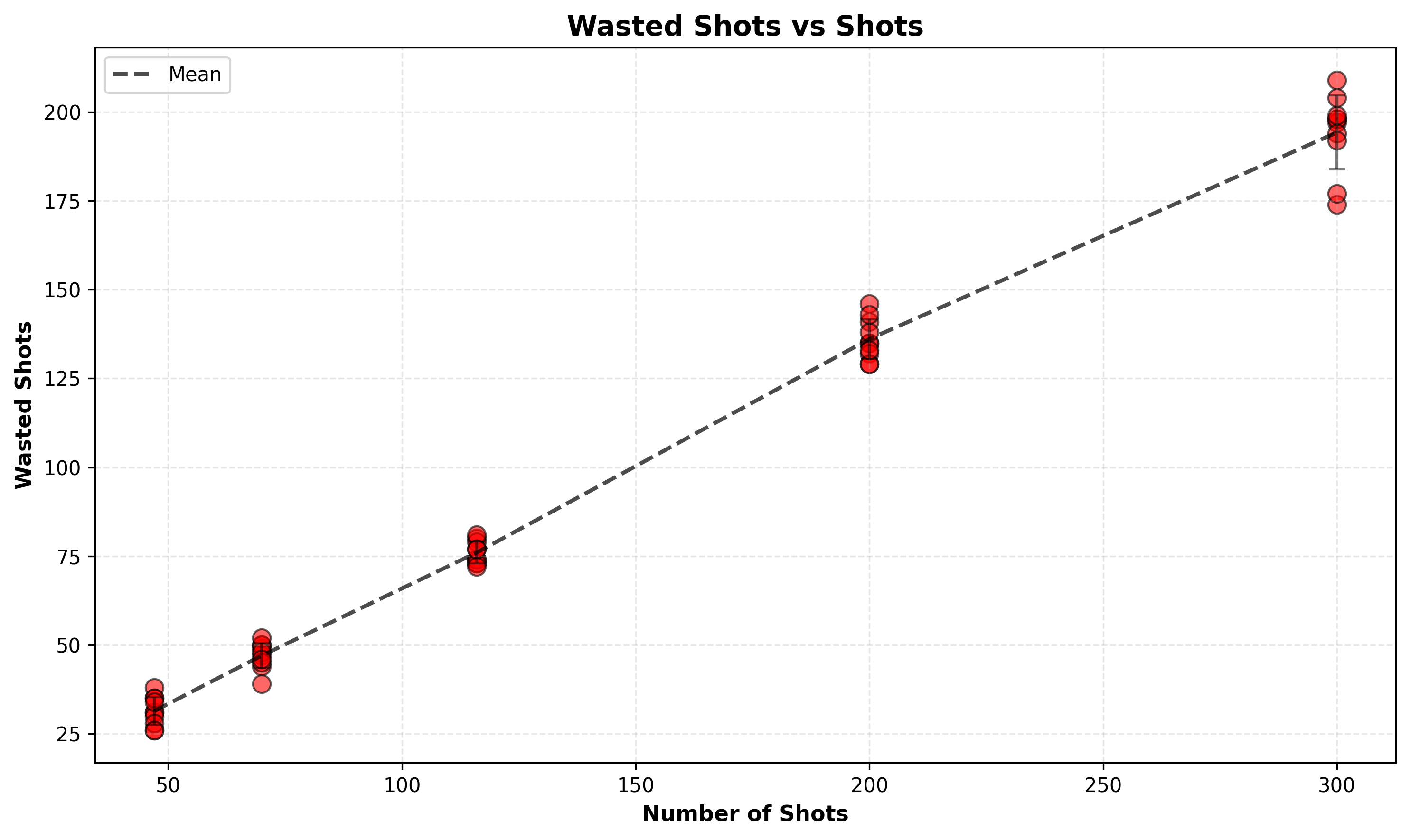}
\caption{Noisy simulator}
\label{fig:wasted_vs_shots:sim}
\end{subfigure}
\caption{Wasted shots versus total shot count. Error bars show standard deviation across 10 experiments per shot count. The relatively constant wasted shot fraction indicates that noise sources scale proportionally with measurement budget.}
\label{fig:wasted_vs_shots}
\end{figure}

All shots that lead to empty voxels or invalid indices are also referred to as \emph{wasted shots}. In \cref{fig:wasted_vs_shots}, we present the overhead of such wasted shots in relation to the total number of shots. Both on hardware and simulator, wasted shots exhibit a positive correlation with total shots. Mean values on hardware range from $35$ wasted shots at $S = 47$ (corresponding to $74$\% wasted-shot fraction) to $214$ wasted shots at $S = 300$ ($71$\% wasted-shot fraction). On the simulator, the corresponding range is from $31$ wasted shots at $S = 47$ ($67$\% wasted-shot fraction) to $194$ wasted shots at $S = 300$ ($65$\% wasted-shot fraction). In both cases, the wasted fraction remains approximately constant across the explored shot range.

\subsection{Reconstruction Recall} \label{subsec:recall}

To quantify the quality at which a given quantum state $\ket{\psi_{M^*}}$ has been decoded with a sequence of measurements, we introduce the
\textit{reconstruction recall}
\begin{equation} \label{eq:reconstruction_recall}
    R := \frac{|\mathcal{A}_{\text{found}}|}{A},
\end{equation}
where $\mathcal{A}_{\text{found}} \subseteq \{1, \ldots, A\}$ denotes the set of atom indices that have been successfully measured at least once across all shots. In an ideal case, a complete reconstruction with $R = 1$ is achieved, which means that all $A$ atoms are successfully detected. By construction, $R$ is a \emph{recall} metric: it measures the fraction of true atoms recovered and does not penalize false positives, such as empty-voxel detections or invalid indices. In other words, a run with $R = 1$ may still contain many wasted shots. The complementary effect of this noise is quantified separately by the \gls{SNR} introduced later in this section.

\begin{figure}[ht!]
\centering
\begin{subfigure}{\columnwidth}
\includegraphics[width=\columnwidth]{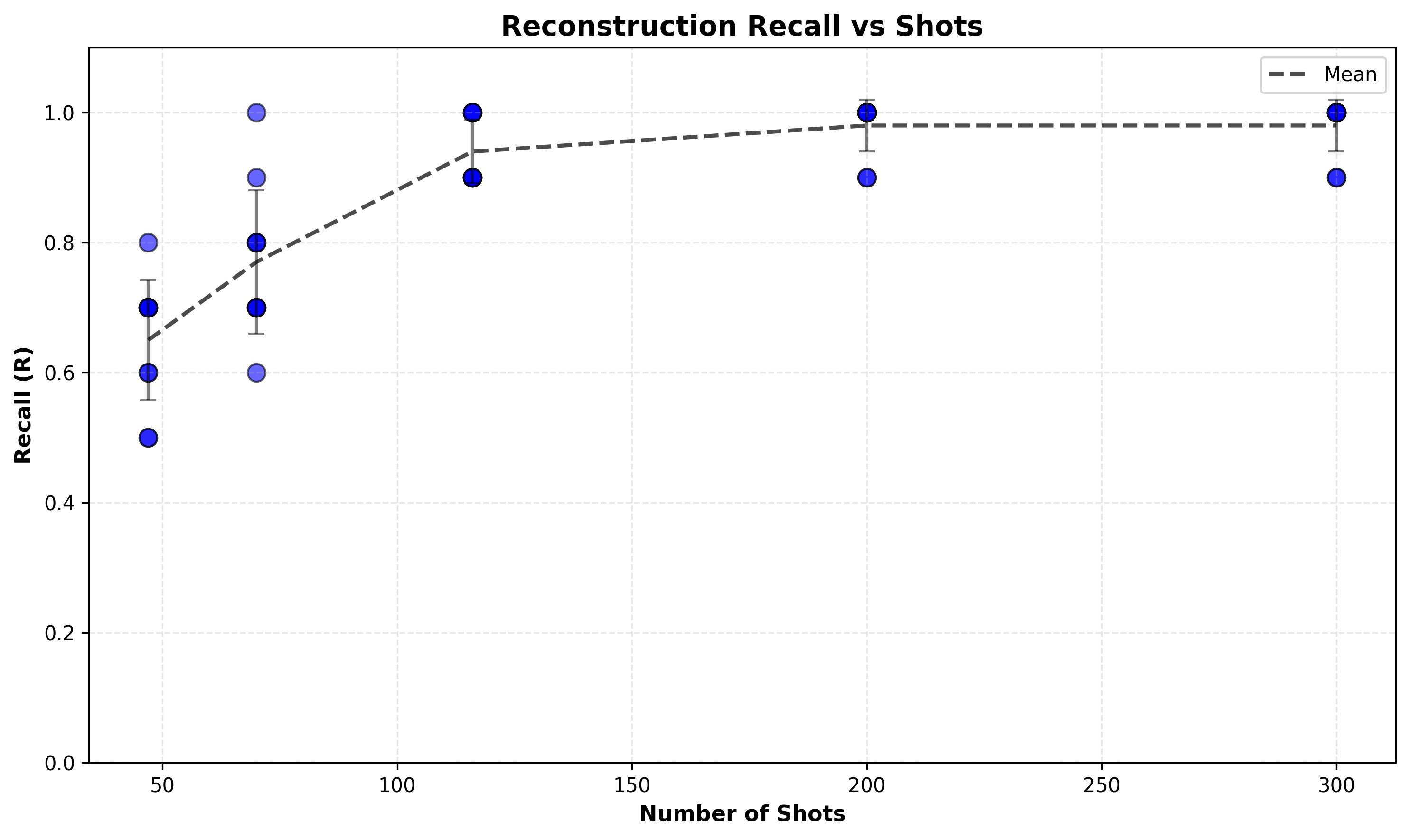}
\caption{\emph{IBM Kingston}}
\label{fig:recall_vs_shots:hw}
\end{subfigure}
\\[.2cm]
\begin{subfigure}{\columnwidth}
\includegraphics[width=\columnwidth]{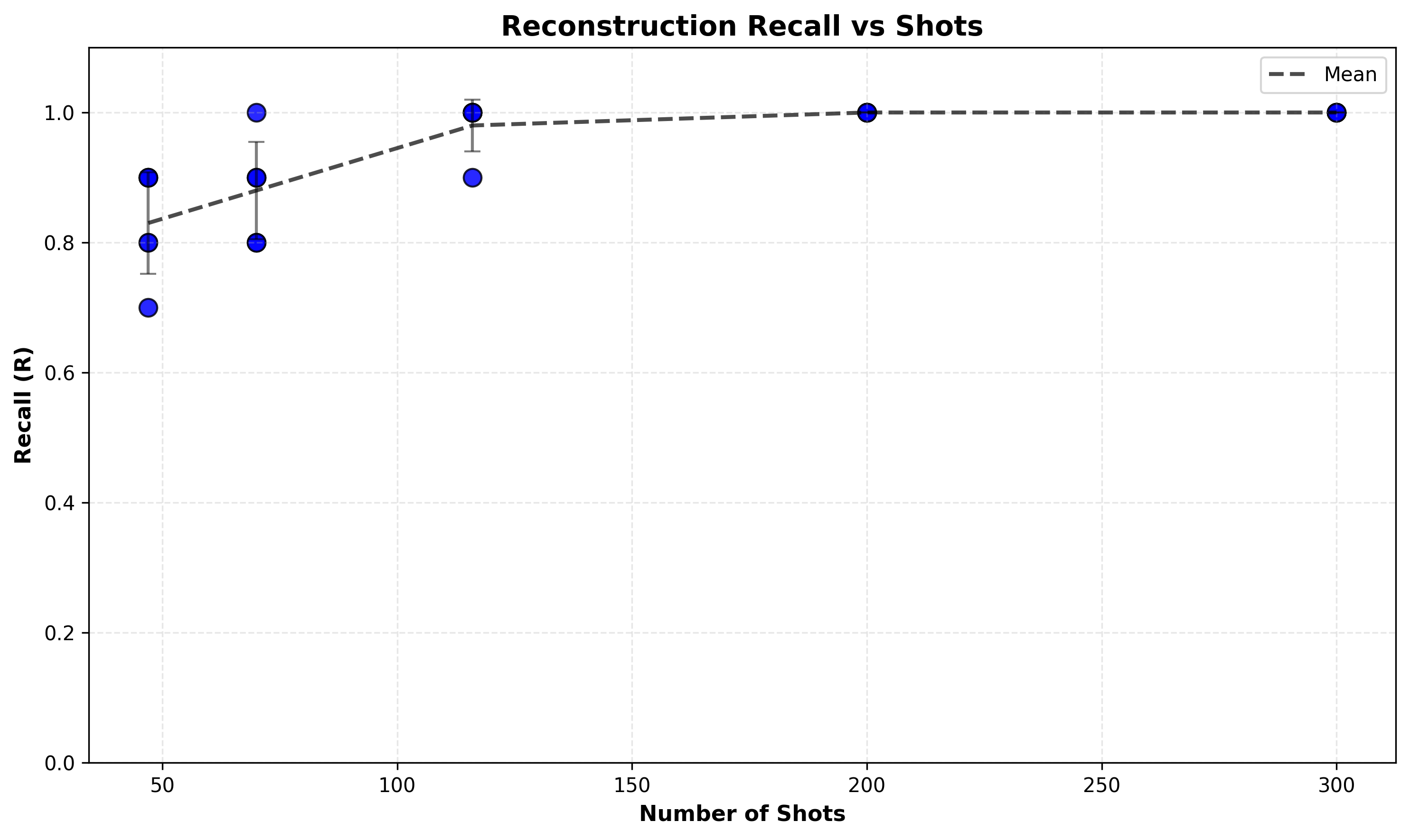}
\caption{Noisy simulator}
\label{fig:recall_vs_shots:sim}
\end{subfigure}
\caption{Reconstruction recall $R$, \cref{eq:reconstruction_recall}, versus number of shots. Each point shows mean recall $\pm$ standard deviation across 10 experiments. The recall exhibits a clear positive trend with increasing shot count, approaching $R = 1$ at higher shot budgets despite hardware noise.}
\label{fig:recall_vs_shots}
\end{figure}

In \cref{fig:recall_vs_shots}, we present the reconstruction recall $R$ as a function of shot count. Each data point represents the mean recall over 10 independent experiments at that shot count, with error bars indicating one standard deviation. Within the shot range explored (47--300), both hardware and simulator demonstrate that recall improvement with shot count follows the expected behavior from coupon collector statistics: higher measurement budgets increase the probability of observing all atoms at least once. On hardware, the mean reconstruction recall $\bar{R}$ exhibits a clear positive correlation with shot count, achieving $\bar{R} = 0.65 \pm 0.09$ at $S = 47$ shots, $\bar{R} = 0.94 \pm 0.05$ at $S = 116$ shots (4 out of 10 experiments reaching $R = 1$), and $\bar{R} = 0.98 \pm 0.04$ at $S = 200$ shots (8 out of 10 reaching $R = 1$). Hence, the target $R = 1$ can be achieved on the hardware with shot counts on the order of $\mathcal{O}(10^2)$, two orders of magnitude fewer than the usual state tomography techniques~\cite{nielsen2010, Stricker_2022, Zambrano2024tQST} while maintaining full molecular recovery within the voxel grid resolution. The simulator achieves higher recall at lower shot counts: $\bar{R} = 0.83 \pm 0.08$ at $S = 47$, $\bar{R} = 0.98 \pm 0.04$ at $S = 116$ (8 out of 10 reaching $R = 1$), and $\bar{R} = 1.00$ at $S = 200$ (all 10 experiments reaching $R = 1$). The expectation value of the reconstruction recall and its connection to the hardware noise is briefly outlined in \cref{app:shots_calculation}.

The reconstruction recall, \cref{eq:reconstruction_recall}, requires prior knowledge of which basis states correspond to atoms, a ground truth that may not be available in a generative setting. In this case, the identity of valid atomic indices must be inferred purely from measurement statistics: basis states corresponding to actual atoms are expected to appear with systematically higher counts than noise-induced empty voxel detections, since all $A$ atoms share equal probability $1/A$ under the ideal equal superposition, while noise is spread across the remaining $C - A$ unoccupied states. We therefore introduce the \gls{SNR} as a ground-truth-free proxy for reconstruction quality, which only requires knowledge about the molecular support size:
\begin{equation}
    \text{SNR} = \frac{\langle N_{\text{valid}} \rangle_{10 \text{ atoms}}}{\langle N_{\text{empty}} \rangle_{182 \text{ voxels}}},
\end{equation}
where $\langle N_{\text{valid}} \rangle$ and $\langle N_{\text{empty}} \rangle$ denote the mean count per valid atomic index and per empty voxel index, respectively. A higher $\text{SNR}$ indicates that atomic basis states are more distinguishable from noise, enabling reliable atom identification without prior knowledge of the molecular structure. Hardware yields $\text{SNR} \approx 7.8$ and the simulator yields $\text{SNR} \approx 10.2$, reflecting the lower noise level of the simulator relative to real hardware. Both values benefit from the larger 192-dimensional encoding space, which spreads noise across 182 empty voxel states.

\subsection{Runtime}
\label{subsec:qpu_time}

Across all experiments, we measure the hardware execution time via execution spans reported by the IBM Runtime primitive, which provide sub-second precision independent of billing granularity. For our 8-qubit, $\sim$1213--1240 gate circuits, the measured runtime is \SI[separate-uncertainty=true]{0.95 \pm 0.41}{\second} (mean $\pm$ standard deviation across 50 experiments), consistent with expectations for circuits of this depth given typical gate times ($\sim$\SI{35}{\nano\second} for single-qubit gates, $\sim$\SI{84}{\nano\second} for two-qubit (CX) gates) and measurement durations ($\sim$\SI{1}{\micro\second}). The IBM wall-clock time (from job start to completion) averages $\sim$\SI{4.5}{\second} per experiment, with the difference ($\sim$\SI{3.5}{\second}) attributable to classical control overhead on the IBM side, including payload loading into control electronics. The IBM billing unit (quantum seconds, ceiling-rounded) is one second per experiment for all shot counts tested, confirming that the chip execution time remains below one second across the 47--300 shot range.

\section{Conclusions} \label{sec:conclusion}

We have presented quantum voxelization encoding, a practical and measurement-efficient strategy for representing molecular structures with discretized spatial information on \gls{NISQ} devices. By discretizing molecular space into a 3D grid and encoding atomic positions and types as equal superpositions of computational basis states, our approach achieves complete reconstruction of a molecule with $A$ atoms with only $\mathcal{O}(A \log A)$ shots. Experimental validation on the 156-qubit \textit{IBM Kingston} processor demonstrates mean reconstruction recall $\bar{R}=0.94{\pm}0.05$ at $S=116$ shots and $\bar{R}=0.98{\pm}0.04$ at $S=200$ shots (8 out of 10 experiments achieving complete reconstruction) for a 10-atom molecule, confirming the practical viability. The proposed encoding may be particularly well-suited for quantum generative modeling architectures, including \gls{QAE}~\cite{romero2017quantum} and \gls{QGAN}~\cite{lloyd2018quantum}, where an efficient measurement scheme is key.

We emphasize that this result concerns only the \emph{readout} (decoding) stage but does not by itself make voxelization an efficient full quantum data-loading procedure. State preparation still requires a comparatively deep circuit ($\sim$1213 gates for our 8-qubit instance), which leads to a scaling bottleneck not addressed by our work.

There are two fundamental challenges associated with the practical use of voxelization encoding. First, identifying an appropriate voxel grid is non-trivial. The finite grid resolution $V$ introduces a quantization error in the representation of atomic positions, bounded by $\frac{\sqrt{3}}{2}\, s_{\text{voxel}}$ per atom (see \cref{subsec:quantization_error}). For applications that require sub-ångström accuracy (e.g., conformational analysis or docking simulations), this spatial discretization may therefore be insufficient. As a result, a fundamental trade-off arises between spatial resolution and the required number of basis states. Second, physically equivalent molecules should ideally be mapped to equivalent quantum states; however, voxel grids do not preserve rotational or translational invariance. Consequently, the choice of the Euclidean coordinate system and the voxel size has a significant influence on the resulting encoding, which can be considered a disadvantage for machine learning tasks that aim to learn underlying molecular physics. Comprehensive scalability studies across molecular datasets (QM9 \cite{ramakrishnan2014quantum}, ZINC \cite{Irwin2005}) with 10--50 atoms may be necessary to characterize the voxel resolution trade-offs and qubit requirements for diverse chemical structures.

There are various directions to extend voxelization encoding in future work. First, by studying different voxelization strategies beyond a Euclidean grid. For example, adaptive or hierarchical voxelization strategies may mitigate the trade-off between spatial resolution and basis-state count by allocating higher resolution only to chemically active regions. Similarly, hybrid approaches combining voxelization for coarse structure with continuous refinement for critical regions could be considered to address the limitation of discretized atom positions. Furthermore, \emph{spherical voxelization} (i.e., discretizing space in radial, azimuthal, and polar coordinates centered at the molecular centroid) would provide natural rotational symmetry about the molecular center. More generally, incorporating molecular graph symmetries or employing equivariant quantum circuit architectures~\cite{meyer2023exploiting, nguyen2022theory} could enable symmetry-preserving encodings that treat rotationally or translationally equivalent conformations identically.

Second, the performance of the encoding could be improved by a more advanced implementation. For example, by enhancing state preparation methods that leverage structured decompositions to reduce circuit depth or by integrating quantum error mitigation techniques (including zero-noise extrapolation and dynamical decoupling).

Finally, integration of voxelization encoding into hybrid quantum-classical generative workflows such as \gls{QAE} and \gls{QGAN} may offer a promising pathway toward practical quantum-assisted molecular design on near-term devices. As a possible pipeline for a generative model, we suggest that a parameterized circuit $U(\theta)\ket{0}^{\otimes n}$ could be trained to produce a sparse distribution over atom voxel/type basis states, approximating a support state of the form of \cref{eq:quantum_state}. Repeated measurements in the computational basis would then yield candidate atoms through the coupon-collector sampling, which can be evaluated with classical post-processing to estimate the occupied support as described in the signal-to-noise analysis in \cref{subsec:recall}. Additionally, chemical-validity filters or bond-inference heuristics could be used to reconstruct a molecular graph from the recovered atom-type voxel-grid.

%% file: appendix.tex
\section*{Appendices}


\section{Molecule Configuration Count} \label{app:molconf}

The configuration set $\hat{\mathcal{C}}$ from \cref{subsec:voxelization_framework} comprises all $\hat{C}$ collision-free voxelized configurations, taking differently ordered representations into account. The number of (unordered) distinguishable configurations for arbitrary molecule compositions is given by
\begin{equation}
\label{eq:hatCp}
    \hat{C}':=\binom{V^3}{A}T^A,
\end{equation}
with $\hat{C}' \leq \hat{C}$; the two counts are related by $\hat{C} = A!\,\hat{C}'$, since $\hat{C}$ counts ordered $A$-tuples whereas $\hat{C}'$ counts unordered sets. For a fixed composition, the number of possible configurations is further reduced. If we consider a molecule with $m_{\tau} \in \mathbb{N}$ atoms of atom type $\tau \in \mathcal{T}$ and $\sum_{\tau \in \mathcal{T}} m_{\tau}=A$, the resulting configuration set is restricted to
\begin{equation}
\label{eq:hatCpp}
    \hat{C}'' := \binom{V^3}{m_1,\dots,m_{T},V^3-A}
\end{equation}
possible configurations with $\hat{C}'' \leq \hat{C}'$. We emphasize that these reductions concern the cardinality of the \emph{molecular} configuration space (i.e., how many distinct molecules the scheme could represent); they do not affect the per-molecule qubit count of our encoding, which is governed by the single-atom configuration space $C=V^3T$.


\section{Preprocessing Pipeline} \label{app:preprocessing}

In \cref{sec:experiments}, we consider the voxelization encoding of ethylamine. The molecular structure undergoes standard cheminformatics preprocessing: SMILES~\cite{doi:10.1021/ci00057a005} is used to specify the molecule. Ethylamine is encoded as \texttt{CCN}, where each character represents an atom (\texttt{C} for carbon, \texttt{N} for nitrogen) and bonds are implicit in the sequence. This SMILES string is then converted to a molecular graph via RDKit, followed by explicit hydrogen addition, 3D embedding via ETKDG~\cite{riniker2015better}, MMFF force field optimization~\cite{halgren1996merck}, and coordinate normalization. The molecular preprocessing pipeline consists of three stages: 3D Embedding, Geometry Optimization, and Coordinate Normalization.

\subsection{3D Embedding}
We use the ETKDG conformer generation algorithm~\cite{riniker2015better} with fixed random seed for reproducibility. The ETKDG (Experimental-Torsion Knowledge Distance Geometry) method generates initial 3D coordinates by:
\begin{enumerate}
\item Computing distance bounds from molecular graph topology
\item Sampling torsion angles from experimental distributions
\item Optimizing the embedding to satisfy geometric constraints
\end{enumerate}

\subsection{Geometry Optimization}
MMFF (Merck Molecular Force Field)~\cite{halgren1996merck} minimization to obtain chemically reasonable bond lengths and angles. The optimization minimizes the molecular energy:
\begin{equation}
E_{\text{total}} = E_{\text{bond}} + E_{\text{angle}} + E_{\text{torsion}} + E_{\text{vdW}} + E_{\text{elec}}
\end{equation}
converging when the gradient norm falls below \SI{e-4}{\kilo\calorie\per\mole\per\angstrom}.

\subsection{Coordinate Normalization}
Centering at the molecular centroid (no rescaling is applied; atomic coordinates retain their physical \si{\angstrom} scale):
\begin{enumerate}
\item Compute centroid: $\mathbf{c} = \frac{1}{A}\sum_{a=1}^{A} \mathbf{r}_a$
\item Center coordinates: $\mathbf{r}'_a = \mathbf{r}_a - \mathbf{c}$
\end{enumerate}
The centered coordinates are used directly for voxelization (\cref{subsec:voxelization_framework}); the grid parameters $V$ and $s_{\text{voxel}}$ are chosen such that all atoms fall within the grid.

This pipeline produces optimized 3D coordinates for all 10 atoms (2 \ce{C}, 1 \ce{N}, 7 \ce{H}). The resulting bond lengths are set by the force field: \ce{C-C} (\SI{1.52}{\angstrom}), \ce{C-N} (\SI{1.46}{\angstrom}), \ce{C-H} (\SI{1.09}{\angstrom}). These reproduce the typical single-bond values from the literature (\SIlist{1.54;1.47;1.09}{\angstrom}) to within a few picometers.




\section{Shot Calculation Methodology} \label{app:shots_calculation}

As described in \cref{subsec:measurement_reconstruction}, the molecular reconstruction problem reduces to the classical \textit{coupon collector problem}~\cite{alma9910292682105765}, where coupons correspond to atoms and drawing a coupon corresponds to performing a shot.

In the following, we first derive confidence bounds for the ideal (noise-free) case in which each shot reveals one of the atoms with uniform probability. Subsequently, we consider the noisy case in which the coupon probabilities are non-uniform and a \emph{wasted} coupon can be drawn, which corresponds to measuring a wasted shot. Both bounds are then evaluated for the ethylamine molecule from \cref{sec:experiments}. Finally, we briefly discuss the expectation value of the reconstruction recall $R$ from \cref{subsec:recall}.

\subsection{Ideal Coupon Collector}

Let $S$ be the number of trials needed to observe all $A$ coupons at least once. The expectation value of $S$ is then given by \cref{eq:coupon_expected} and scales as $\mathcal{O}(A\log A)$, whereas the corresponding variance
\begin{equation}
    \text{Var}[S] = A^2 \sum_{i=1}^{A} \frac{1}{i^2} - A\,H_A \approx \frac{\pi^2}{6}\,A^2,
\end{equation}
grows quadratically in $A$.

After $s$ draws, the probability that a fixed coupon has not been observed is $\left(1-\frac{1}{A}\right)^s$. By applying a union bound over all $A$ coupons, the probability of incomplete reconstruction is bounded by 
\begin{equation}
    \Pr[S > s] \leq A\left(1-\frac{1}{A}\right)^s,
\end{equation}
hence
\begin{equation}
    \Pr[S > s] \leq A e^{-s/A}
\end{equation}
due to the standard inequality $1-x \leq e^{-x} \,\forall\, x \in \mathbb{R}$. Choosing $s := A(\ln A + t)$ yields $\Pr[S > s] \leq e^{-t}$. Furthermore, setting the failure probability $\epsilon := e^{-t}$ yields $t = \ln(1/\epsilon)$ and results in \cref{eq:shots_confidence}. Thus, for a molecule with $A$ atoms and desired confidence level $1-\epsilon$, the required number of shots is given by $S_{\epsilon}$.

\subsection{Noisy Coupon Collector}

In the noisy case, the probabilities can be non-uniform and there is a nonzero probability of drawing a non-informative outcome in form of a wasted coupon. Let $p_i$ denote the probability to observe the $i$th coupon, then the probability of a useful draw is given by
\begin{equation}
    \eta := \sum_{i=1}^{A} p_i.
\end{equation}
The remaining probability mass $1-\eta$ corresponds to drawing the wasted coupon. The probability to miss at least one coupon after $s$ draws is given by
\begin{equation}
    \Pr[S > s] \leq \sum_{i=1}^{A} (1-p_i)^s \leq \sum_{i=1}^{A} e^{-s p_i} \leq \sum_{i=1}^{A} e^{-s p_{\mathrm{min}}}
\end{equation}
with
\begin{equation}
    p_{\mathrm{min}} := \min_{i} p_i.
\end{equation}
Assuming $p_{\mathrm{min}}>0$, imposing
$A e^{-s p_{\mathrm{min}}}\leq \epsilon$ yields the sufficient shot bound
\begin{equation}
    S_{\epsilon,p_{\mathrm{min}}} \geq \frac{\ln A + \ln \frac{1}{\epsilon}}{p_{\mathrm{min}}}.
\end{equation}

If noise acts approximately as erasure, so that useful draws are still
uniformly distributed over the $A$ coupons, then $p_i=\eta/A$ and the
general bound reduces to
\begin{equation} 
    S_{\epsilon,\eta} \approx \frac{A}{\eta} \left( \ln A + \ln \frac{1}{\epsilon} \right),
\end{equation}
which corresponds to \cref{eq:shots_confidence} with an additional factor $1/\eta$ and consequently $S_{\epsilon,\eta=1}=S_{\epsilon}$. Under these conditions, hardware noise increases the ideal shot budget by approximately $1/\eta$, with further degradation when the atom probabilities are non-uniform. According to the measurement statistics from \cref{fig:noise_composition}, we can estimate $\eta \approx 0.29$. However, \cref{fig:valid_vs_empty} clearly indicates a non-uniform distribution $p_i \neq \eta/A$.

\subsection{Ethylamine Example}

For ethylamine with $A=10$ atoms, we obtain the shot budgets listed in \cref{tab:shot_budgets}. All entries are rounded up to the nearest integer, since the number of shots
must be an integer.

\begin{table*}[tb]
\centering
\begin{tabular}{cccccc}
\toprule
Confidence & $S_{\epsilon}$ & $S_{\epsilon,\eta} (\eta=0.8)$ & $S_{\epsilon,\eta} (\eta=0.6)$ & $S_{\epsilon,\eta} (\eta=0.4)$ & $S_{\epsilon,\eta} (\eta=0.2)$ \\
\midrule
$90.00\%$ & 47 & 58 & 77 & 116  & 231  \\
$95.00\%$ & 53 & 67 & 89 & 133  & 265  \\
$97.50\%$ & 60 & 75 & 100 & 150 & 300  \\
$99.00\%$ & 70 & 87 & 116 & 173 & 346  \\
$99.90\%$ & 93 & 116 & 154 & 231 & 461 \\
$99.99\%$ & 116 & 144 & 192 & 288 & 576 \\
\bottomrule
\end{tabular}
\caption{Shot budgets for ethylamine with $A=10$ atoms.} \label{tab:shot_budgets}
\end{table*}


\subsection{Reconstruction Recall}

Presuming a uniform atom measurement with eraser noise $p_i=\eta/A$, the expected reconstruction recall after $S$ shots, \cref{eq:reconstruction_recall}, reads
\begin{equation} \label{eq:reconstruction_recall_expected}
    \mathbb{E}[R_S] = 1 - \left( 1 - \frac{\eta}{A} \right)^S,
\end{equation}
which allows to determine
\begin{equation}
    \eta = A \left[ 1 - \left( 1 - \mathbb{E}[R_S] \right)^{1/S} \right].
\end{equation}
An estimator for $\eta$ can therefore be based on the mean reconstruction recall $\bar{R}$ with
\begin{equation}
    \hat{\eta} := A \left[ 1 - \left( 1 - \bar{R} \right)^{1/S} \right].
\end{equation}
According to the reported results from \cref{subsec:recall}, we find $\hat{\eta} = 0.22$ for $\bar{R} = 0.65$ at $S = 47$, $\hat{\eta} = 0.24$ for $\bar{R} = 0.94$ at $S = 116$, and $\hat{\eta} = 0.19$ for $\bar{R} = 0.98$ at $S = 200$.


\section{Device Properties} \label{app:device_properties}

For the numerical experiments in \cref{sec:experiments}, we use the \emph{IBM Kingston} device, a 156-qubit superconducting transmon processor with heavy-hexagon topology. Device characteristics from calibration data (April 2026) are summarized below.

\subsection{Coherence Times}
\begin{itemize}
\item \textbf{$T_1$ (energy relaxation):} Ranges from tens to $\approx \SI{400}{\micro\second}$ across qubits
\item \textbf{$T_2$ (dephasing):} Ranges from tens to $\approx \SI{200}{\micro\second}$ across qubits
\item \textbf{Typical values:} Mean $T_1 \approx \SI{150}{\micro\second}$, mean $T_2 \approx \SI{100}{\micro\second}$
\end{itemize}

\subsection{Gate Performance}
\begin{itemize}
\item \textbf{Single-qubit gates (RZ, SX):} Error rates $< 0.1\%$, duration $\sim$\SI{35}{\nano\second}
\item \textbf{Two-qubit gates (CX):} Error rates predominantly below 3\%, median duration $\sim$\SI{84}{\nano\second}
\item \textbf{Gate count:} Transpiled circuits contain $\sim$1213 gates
\end{itemize}

\subsection{Readout Fidelity}
\begin{itemize}
\item \textbf{Majority of qubits:} Readout error $< 5\%$
\item \textbf{Some qubits:} Substantially higher error rates (up to 10-15\%)
\item \textbf{Assignment fidelity:} Typically $F_{|0\rangle} > 0.95$, $F_{|1\rangle} > 0.90$
\end{itemize}

\subsection{Topology}
Heavy-hexagon architecture with:
\begin{itemize}
\item 156 physical qubits
\item Connectivity degree: 2-3 neighbors per qubit
\item Native gates: \texttt{id, rz, sx, x, cx}
\end{itemize}


\section{Transpilation} \label{app:transpilation}

Since quantum hardware operates on a fixed native gate set and a constrained qubit connectivity graph, an abstract circuit must undergo \textit{transpilation}, a compilation step that rewrites the circuit into hardware-compatible gates and routes two-qubit operations through physically connected qubits at the cost of additional gates. The resulting \textit{physical circuit} is the hardware-executable version of the abstract encoding circuit; its depth reflects both the complexity of state preparation and the overhead introduced by this compilation step. 

For the numerical experiments in \cref{sec:experiments}, transpilation to \emph{IBM Kingston} proceeds through the following stages:

\begin{enumerate}
\item \textbf{Gate decomposition:} Abstract gates decomposed into native basis \{\texttt{rz, sx, cx}\}.
\item \textbf{Qubit layout:} 8 logical qubits mapped to physical qubits on the device using Qiskit's \texttt{SabreLayout} algorithm, which minimizes SWAP gate overhead while respecting device connectivity constraints.
\item \textbf{Routing:} CNOT gates between non-adjacent qubits are routed through intermediate qubits via SWAP insertion.
\item \textbf{Gate optimization:} Consecutive single-qubit rotations are merged, and commuting gates are reordered to reduce depth.
\end{enumerate}

We select \texttt{optimization\_level=3} to maximize gate reduction and circuit simplification, which is critical for minimizing decoherence effects on \gls{NISQ} hardware.

Since \texttt{SabreLayout} is a stochastic algorithm, different runs can produce different qubit assignments and SWAP overhead, leading to varying circuit depths. To mitigate this, a pre-computation step was performed prior to the experiments: the circuit was transpiled multiple times and the resulting transpiled circuit with the shortest depth was retained. This fixed circuit was then used for all 50 experiments, yielding a consistent depth of $1213$ gates across every run.

The resulting circuit depth represents a significant increase over the abstract circuit due to: (i) basis gate decomposition, (ii) SWAP insertion for routing non-adjacent CNOT gates, and (iii) optimization passes. This depth approaches the upper limit of coherence-limited computational budgets for superconducting qubits ($\sim$50--100 gates for high-fidelity operations). However, our experimental results demonstrate that near-complete molecular reconstruction ($\bar{R}=0.98$ at 200 shots) remains achievable despite this circuit complexity, owing to the robustness of equal superposition encoding: the measurement-based reconstruction tolerates moderate gate errors as long as the correct basis states retain sufficient probability mass to be sampled within the allocated shot budget.